# An Energy-Efficient FPGA-based Deconvolutional Neural Networks Accelerator for Single Image Super-Resolution

Jung-Woo Chang, Keon-Woo Kang, and Suk-Ju Kang, *Member, IEEE*

*Abstract*—Convolutional neural networks (CNNs) demonstrate excellent performance in various computer vision applications. In recent years, FPGA-based CNN accelerators have been proposed for optimizing performance and power efficiency. Most accelerators are designed for object detection and recognition algorithms that are performed on low-resolution (LR) images. However, real-time image super-resolution (SR) cannot be implemented on a typical accelerator because of the long execution cycles required to generate high-resolution (HR) images, such as those used in ultra-high-definition (UHD) systems. In this paper, we propose a novel CNN accelerator with efficient parallelization methods for SR applications. First, we propose a new methodology for optimizing the deconvolutional neural networks (DCNNs) used for increasing feature maps. Secondly, we propose a novel method to optimize CNN dataflow so that the SR algorithm can be driven at low power in display applications. Finally, we quantize and compress a DCNN-based SR algorithm into an optimal model for efficient inference using on-chip memory. We present an energy-efficient architecture for SR and validate our architecture on a mobile panel with quad-high-definition (QHD) resolution. Our experimental results show that, with the same hardware resources, the proposed DCNN accelerator achieves a throughput up to 108 times greater than that of a conventional DCNN accelerator. In addition, our SR system achieves an energy efficiency of 144.9 GOPS/W, 293.0 GOPS/W, and 500.2 GOPS/W at SR scale factors of 2, 3, and 4, respectively. Furthermore, we demonstrate that our system can restore HR images to a high quality while greatly reducing the data bit-width and the number of parameters compared to conventional SR algorithms.

## I. INTRODUCTION

RECENTLY, object detection [1]–[3], recognition [4]–[6], and natural language processing [7] have attracted considerable attention because of the emergence of convolutional neural networks (CNNs). As a result, extensive studies on CNN accelerators have been conducted in order to implement CNN algorithms in real-time systems. In particular, in hardware implementation, FPGA-based CNN accelerators are more energy-efficient than those based on GPUs and can perform more massive parallel processing than those based on CPUs [8]. In addition, compared to ASICs, FPGAs are more flexible enough for handling the rapid evolution of CNNs [9], [10].

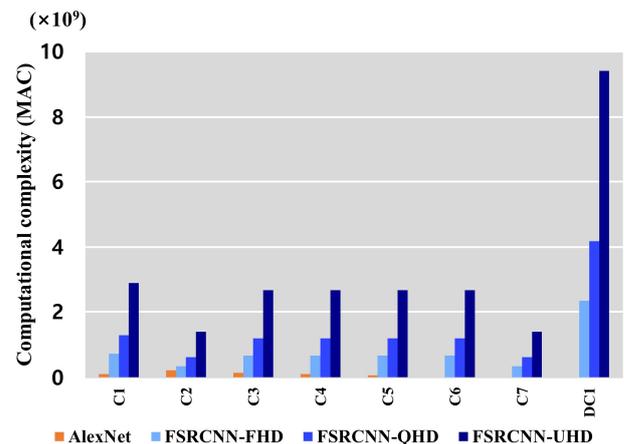

Fig. 1. Comparison of the computational complexity between AlexNet composed of five convolutional layers and FSRCNN composed of seven convolutional layers and one deconvolutional layer, where the convolutional layers are from C1 to C7, and the deconvolutional layer is DC1.

Most CNN accelerator-related studies [8]–[22] have been focused on object detection and recognition applications. In recent years, research studies on image super-resolution (SR) using CNNs have been attracting considerable attention, because CNN-based methods can reconstruct images with a higher peak signal-to-noise ratio (PSNR) than conventional methods [23]–[25]. However, since most CNN accelerators are designed for object detection and recognition algorithms, the following problems can occur when SR algorithms are implemented in a typical accelerator.

First, SR requires a considerably higher resolution image input than object detection and recognition algorithms to generate full-high-definition (FHD), quad-high-definition (QHD), and ultra-high-definition (UHD) videos for mobile applications or TV services. Fig. 1 shows a computational complexity comparison between AlexNet [4] and FSRCNN [25], which is a well-known deep neural network (DNN)-based SR algorithm. Most object classifiers operate on input images with a pixel resolution of less than 256×256 [4]–[6]. Since the resolution used in object classifiers is less than that used in SR, FSRCNN requires 38.82 times more multiply-accumulate (MAC) operations than AlexNet when generating UHD images.

Secondly, recent DNN-based SR algorithms, including

This research was supported by the MSIT(Ministry of Science and ICT), Korea under the ITRC(Information Technology Research Center) support program(IITP-2018-0-01421) supervised by the IITP(Institute for Information & communications Technology Promotion) and National Research Foundation of Korea(NRF) grant funded by the Korea Government(MSIT) (No. 2018R1D1A1B07048421)

J.-W. Chang, K.-W, Kang, and S.-J. Kang are with the Department of Electronic Engineering, Sogang University, Seoul 121-742, Republic of Korea. S.-J. Kang is the corresponding author (e-mail: sjkang@sogang.ac.kr)

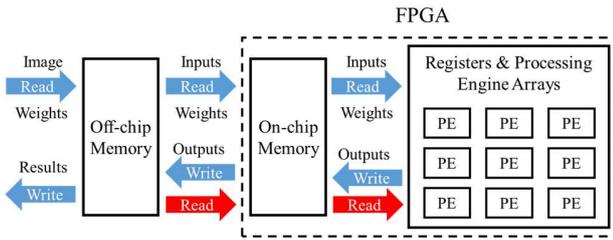

Fig. 2. Three levels of general hardware DCNN accelerator hierarchy: 1) off-chip memory; 2) on-chip memory; 3) processing elements (PEs). Unlike the CNN accelerator, DCNN accelerator has an overhead to read the outputs stored in the off-chip memory due to the overlapping sum problem. In this figure, the red arrows indicate additional read operations.

TABLE I
NOTATIONS OF PARAMETERS

| Notation | Explanation |
|---|---|
| $H_{in}$ / $W_{in}$ | Height/width of input feature maps |
| $H_{out}$ / $W_{out}$ | Height/width of output feature maps |
| $M$ / $N$ | Number of output/input feature maps |
| $K_D \times K_D$ | Kernel size for deconvolutional layer |
| $K_C \times K_C$ | Kernel size for convolutional layer |
| $S$ / $P$ | Stride / number of padding |
| $T_m$ / $T_n$ | Tile size for the number of output/input feature maps |
| $T_k$ | Tile size for kernel size |
| $W$ | Weight coefficient for deconvolutional layer |
| $\mathbf{W_D}$ | Weights of the deconvolutional layer |
| $\mathbf{W_C}$ | Weights of the convolutional layer |

TABLE II
KEY ABBREVIATION DESCRIPTION

| Abbreviation | Description |
|---|---|
| CNN | Convolutional neural network |
| CLP | Convolutional layer processor |
| DCNN | Deconvolutional neural network |
| DCLP | Deconvolutional layer processor |
| DNN | Deep neural network |
| FHD / QHD / UHD | Full/quad/ultra-high-definition |
| GOPS | Giga operations per second |
| HR / LR | High/low-resolution |
| MAC | Multiply-accumulate |
| PE | Processing element |
| PSNR | Peak signal-to-noise ratio |
| SR | Super-resolution |
| TDC | Transforming deconvolutional layer into convolutional layer |

FSRCNN, use deconvolutional neural networks (DCNNs) [26] at the end of the entire network to reconstruct high-resolution (HR) images from low-resolution (LR) images. The deconvolutional layer has the highest computational complexity; it uses a maximum of 6.75 times more MAC operations than convolutional layers, as shown in Fig. 1. Moreover, unlike CNNs, DCNNs create up-scaled output blocks in terms of the kernel size and accumulate pixel values within the output blocks generated from the neighboring pixels.

As shown in Fig. 2, when DCNNs are implemented in hardware, additional operations are required that aren't required in CNNs to load the previously obtained output pixels in the memory, update the pixel values in the processing elements (PEs), and store them in the memory. This is called the overlapping sum problem [27], [28].

The conventional DCNN accelerator [27] attempts to solve this problem by using formulas to locate the positions of the input pixels needed to generate the output pixel through a reverse looping method. A deconvolutional layer processor (DCLP) was designed to perform parallel operations based on the size of tile parameters by applying loop optimization techniques [10] that remove the dependence of loops. However, the reverse looping method, which requires additional loads before each PE, has a large hardware overhead due to limited resources and is not energy efficient. Additionally, the conventional DCNN accelerator does not optimize the high computational complexity of the deconvolutional layer.

There is a method of generating HR images from LR-sized feature maps through a sub-pixel convolutional layer [29]. This layer performs the same operation as the convolutional layer but combines the LR sized output feature maps into a HR image. However, since zero-weights do not exist in convolution filters, a dense CNN accelerator is required. Therefore, the sub-pixel convolutional layer is inefficient in high complexity SR applications.

In general, when implemented with on-chip memory, it is difficult to store all the data required for CNN-based algorithms, except with binarized feature maps [19]; this is because of the large size of the 3D feature maps obtained each time the convolutional layer is processed. Therefore, most FPGA-based CNN accelerators use an off-chip memory and perform off-chip data transfer and computation simultaneously through ping-pong operations. As a result, MAC operations can be performed continuously through a convolutional layer processor (CLP) using loop optimization techniques. Even if ping-pong operations are applied through double buffers, a CLP cannot perform the subsequent operation until a large number of output feature maps are stored in the off-chip memory. In this case, the performance of the accelerators is degraded [17], [18].

In previous papers, CNN fusion architectures were proposed for reducing a large amount of off-chip data transfers [17], [18]. Fusion architectures are designed with various CLPs for processing multiple convolutional layers. Therefore, the data generated after each CLP is operated are transferred to the next layer processor using the on-chip memory. The off-chip data transfer only occurs on the first and last fused layers. However, CNN fusion architectures still require communication with off-chip memory, which is not energy-efficient.

In order to reduce the bandwidth between the accelerator and off-chip memory, Brainwave [30] stores the DNN model and intermediate data in on-chip storage to simplify the off-chip interconnect. This requires quantization and compression of the DNN model. However, Brainwave is limited in that it does not support DCNN.

In this paper, we propose a novel SR-based DNN accelerator for real-time HR image generation with efficient dataflow. The main contributions of this paper are as follows.

- We propose a novel DCNN accelerator that can be massively parallelized by transforming the deconvolutional layer into the convolutional layer (the TDC method). We identified a load imbalance problem during the convolution process executed by the TDC method in our previous work [28]. To overcome this problem, we propose a new load balance-aware TDC method that increases the efficiency of sparse matrix multiplication.
- We propose a dataflow for hardware acceleration to store the intermediate data between the layers using the on-chip memory.
- We quantize and compress a representative DCNN-based SR algorithm, called FSRCNN, into an optimal model for efficient inference using on-chip memory. If we design other SR algorithms, the same optimization process can be done to be implemented in on-chip memory. We present an energy-efficient DNN-based SR system. Our system achieves an energy efficiency of 144.9 GOPS/W, 293.0 GOPS/W, and 500.2 GOPS/W for SR scale factors of 2, 3, and 4, respectively.

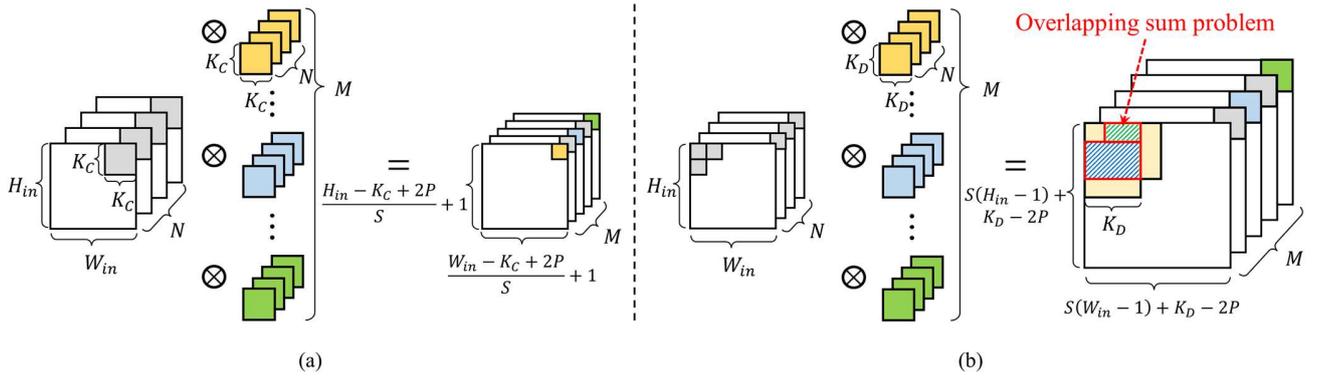

Fig. 3. Comparison of computation schemes for (a) convolutional layer and (b) deconvolutional layer. The convolutional layer extracts the local feature from the input feature maps through trained filters and stores them in the output feature maps. The deconvolutional layer restores the image from input feature maps containing local features through deconvolution. In the deconvolutional layer, there is an overlapping sum problem where adjacent output blocks overlap during the 2D deconvolution operation. The red bounding boxes indicate areas where the overlapping sum problem appears.

The rest of this paper is organized as follows. Section II gives an overview of the CNN and DCNN algorithms. Section III describes the proposed methodology for the DCNN accelerator. Section IV presents the proposed hardware architecture for SR systems and the details of the hardware implementation. Section V presents the experimental results compared to state-of-the-art methods and shows the results of the hardware implementation. Finally, we conclude this paper in Section VI.

For clear description, Table I shows the parameter notations used in the convolutional and deconvolutional layers. Table II shows key abbreviations used in this paper.

## II. BACKGROUND

### A. Convolutional Neural Networks

Fig. 3(a) shows the convolutional layer constituting the CNN structure. The convolutional layer receives input feature maps, which are arranged in three dimensions, $H_{in} \times W_{in} \times N$. Then, it creates output feature maps, which are the results from the input feature maps obtained using learned weights. The process of generating output feature maps is as follows. First, input blocks that move by a defined stride in the input feature maps perform convolution with weights. The kernel size is $K_C \times K_C$, the number of the kernels is $M \times N$, and the stride is $S$. To create $M$ output feature maps, all the $N$ outputs generated by the same type of convolution filter are added together with biases. Finally, the activation function [31] transforms the outputs of the three-dimensional convolution.

### B. Deconvolutional Neural Networks

Fig. 3(b) illustrates the deconvolutional layer that comprises the DCNN. The deconvolutional layer moves the sliding window at stride intervals in the output feature maps rather than in the input feature maps. The output size is $K_D \times K_D$. Therefore, as shown in Fig. 3(b), the overlapping sum problem, where the output blocks are overlapped with the neighboring output blocks, occurs in the green and blue regions. The outputs located in the green region can be easily updated using on-chip buffers because they do not overlap vertically with adjacent output blocks. On the other hand, outputs located in blue region must be called from the memory whenever they overlap with vertically adjacent output blocks. Consequently, it is difficult to store large amounts of intermediate data in on-chip memory to update the previously generated outputs. Unless the final outputs are no longer overlapping with neighboring blocks, the processor must read the output that is already stored in memory and update and store it again. This inefficient dataflow interferes with the ping-pong operation, which can overlap the computation of the processor with the data transfer time [10].

## III. PROPOSED DCNN ACCELERATOR

### A. TDC Method

Each pixel in the input feature map generates an output block of $K_D \times K_D$ through deconvolution. However, there is a problem of overlapping with the output blocks generated from neighboring input pixels. Fig. 4 shows examples where output blocks generated from adjacent inputs overlap with each other. We must add all the overlapping areas every time the input pixels perform 2D deconvolution. To avoid this overlapping sum problem, we must determine the number of input pixels required to generate an output block that no longer overlaps.

Each output block can overlap with adjacent output blocks within a range of $\lfloor K_D/2 \rfloor$, as shown in Fig. 4. Since the value of the stride $S$ is always greater than 1 for up-scaling, the input pixels that are mapped to the output feature map, which are depicted as red bounding boxes, are spaced apart by $S$. Thus, the value $N_O$, which indicates how many horizontal (or vertical) neighboring blocks overlap within $\lfloor K_D/2 \rfloor$ can be calculated as

$$N_O = \left\lfloor \frac{K_D}{2} \right\rfloor \times \frac{1}{S}. \quad (1)$$

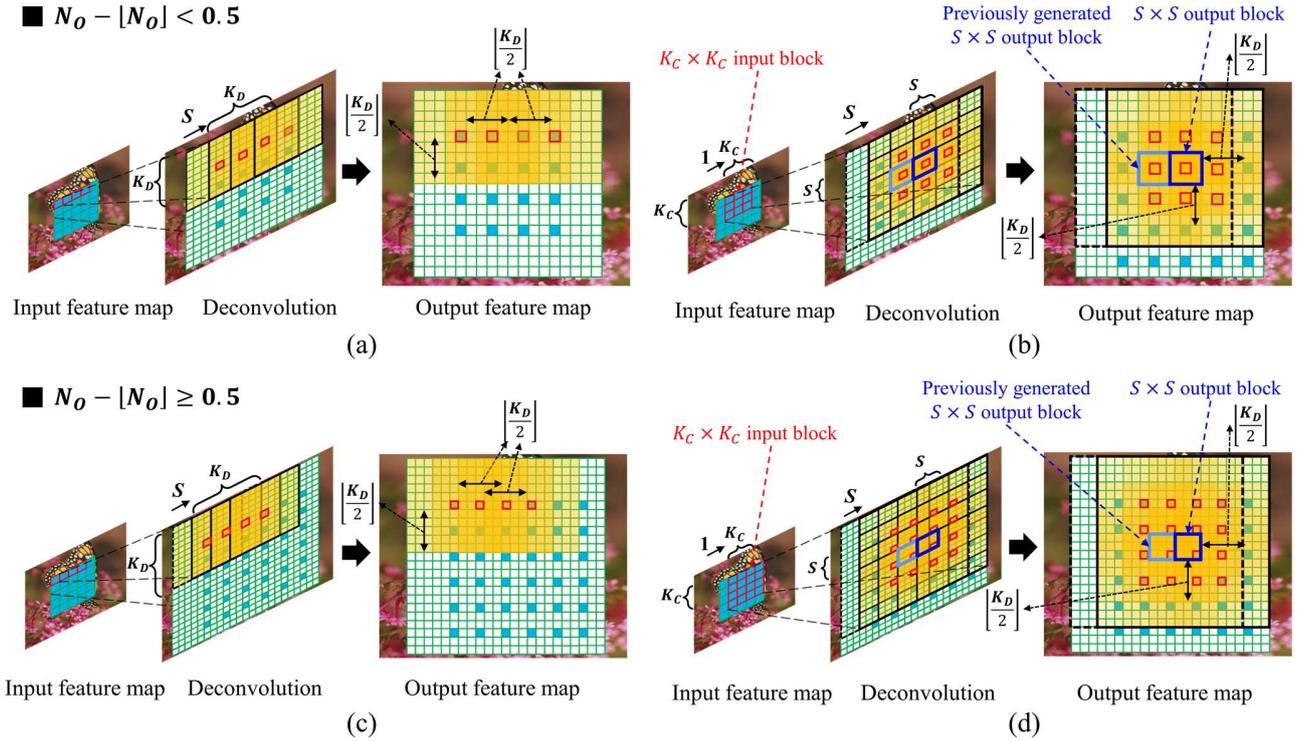

Fig. 4. Examples of overlapping output blocks with adjacent input pixels through deconvolution. To be specific, (a) and (b) are examples when the fractional value of $N_O$ is less than 0.5, and (c) and (d) are larger than 0.5. (a) and (c) illustrate how overlapping area occur in the output feature map through the 2D deconvolution. (b) and (d) show how the $S \times S$ output block can be generated from the $K_C \times K_C$ input block. In this figure, $K_D$ and $S$ are 9 and 3 in (a) and (b), respectively, and 11 and 3 in (c) and (d), respectively. $K_C$ is 3 and 4 in (b) and (d) respectively.

The fractional value of $N_O$ determines how the current output block overlaps with the most distant output block. Fig. 4 shows a comparison of cases where the integer values are the same but the fractional values differ. As shown in Fig. 4(a), when the fractional value of $N_O$ is less than 0.5, the top leftmost output block overlaps two neighboring output blocks within $\lfloor K_D/2 \rfloor$, but does not overlap with the adjacent third output block on the same line. Conversely, if the fractional value is greater than 0.5, the top leftmost output block overlaps all three neighboring output blocks, as shown in Fig. 4(c). Considering both possible cases, the size of the input block $K_C \times K_C$ that can produce non-overlapping output with the adjacent output block can be determined as

$$K_C = \begin{cases} 2 \times \lfloor N_O \rfloor + 1, & \text{if } N_O - \lfloor N_O \rfloor < 0.5 \\ 2 \times \lceil N_O \rceil, & \text{if } N_O - \lfloor N_O \rfloor \geq 0.5 \end{cases}. \quad (2)$$

Using the property that output blocks are spaced apart by $S$ in the deconvolutional layer, the $K_C \times K_C$ input block generates the $S \times S$ output block through the $K_D \times K_D$ deconvolution filters. Figs. 4(b) and (d) show an example of how the $S \times S$ output block can be generated from the $K_C \times K_C$ input block. Depending on the fractional value of $N_O$, input blocks with different sizes slide in the input feature map and generate the blue bounding box in the output feature map, which represents an $S \times S$ output block. The light blue bounding box represents an $S \times S$ output block created in the previous input block.

The computation process of producing each output pixel consists of MAC operations between input pixels and weight coefficients of the deconvolution filter. In this process, there is a new source of massive parallelism because there is no data dependency in creating output pixels. Specifically, each output pixel can be generated from the convolution between the $K_C \times K_C$ input blocks and the convolution filters equal to the size of the input block. As shown in Fig. 5, we apply the new source of the parallelization in the hardware implementation. Through the TDC method, the pixels in the $S \times S$ output block can all be created on the same timeline. Specifically, we convert the spatial domain in the HR output feature map to generate each pixel in the $S \times S$ output block separately in different channels. This increases the number of output feature maps by $S^2$ times. Likewise, we apply the TDC method for all feature maps.

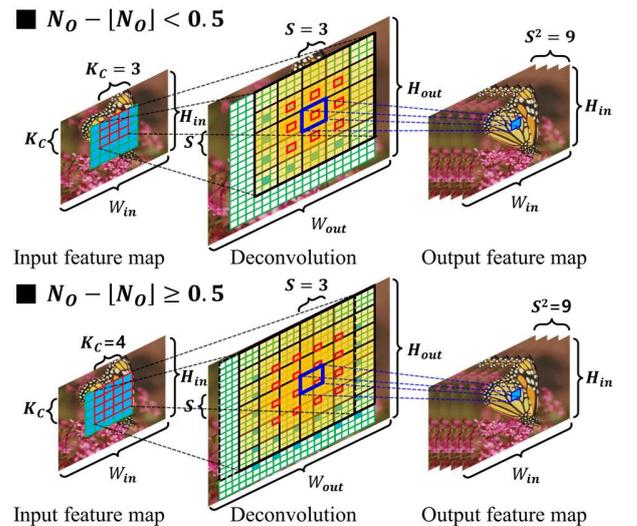

Fig. 5. Visualization of when a new parallelization source is applied at deconvolutional layer according to the fractional value of $N_O$. All the pixels in the $S \times S$ output block can be created simultaneously.

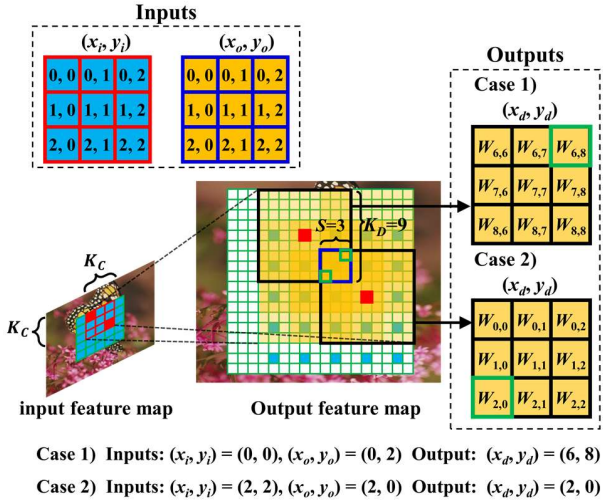

Fig. 6. Example of when inverse coefficient mapping is performed. By the inverse coefficient mapping, we obtain the indices of the weight coefficient when given the input and output pixels respectively located in the red square and the green bounding box. The blue bounding box represents the $S \times S$ output block.

### B. Inverse Coefficient Mapping

We now describe the acquisition of the weights of the newly created convolutional layer through the TDC method. We declare $(x_i, y_i)$, $(x_d, y_d)$, and $(x_o, y_o)$ as the indices of an input pixel, a weight coefficient of a deconvolutional layer, and an output pixel, respectively. The range of each pixel is defined as

$$0 \leq x_i < K_C, \quad 0 \leq y_i < K_C$$
$$0 \leq x_d < K_D, \quad 0 \leq y_d < K_D . \quad (3)$$
$$0 \leq x_o < S, \quad 0 \leq y_o < S$$

For mapping the weights of the deconvolutional layer to those of the convolutional layer, we propose an inverse coefficient mapping to find $(x_d, y_d)$ corresponding to $(x_i, y_i)$. Fig. 6 shows an example of when the inverse coefficient mapping is performed. We must find the indices of the weight coefficient corresponding to the input pixel, which is a red rectangle, to produce an output pixel represented by a green bounding box. The overall process for inverse coefficient mapping is as follows.

As shown in Fig. 7, we divide the inverse coefficient mapping into two processes. First, we obtain the relative position $(x_r, y_r)$ using $(K_D - S \times x_i, K_D - S \times y_i)$. This is because input pixels are shifted by the stride $S$ in the output feature map to produce output blocks. However, since output blocks are created as two types according to the fractional value of $N_O$, their relative position depends on this value. Fig. 7(a) shows that, if the fractional value of $N_O$ is less than 0.5, the relative position is point A. In contrast, Fig. 7(b) shows that, if the fractional value of $N_O$ is greater than 0.5, the relative position is point B. To adjust the relative positions of both cases to the same position, we add 1 if the fractional value of $N_O$ is greater than 0.5. Accordingly, the relative position can be calculated as

$$x_r = \begin{cases} K_D - (S \times x_i), & \text{if } N_O - \lfloor N_O \rfloor < 0.5 \\ K_D + 1 - (S \times x_i), & \text{if } N_O - \lfloor N_O \rfloor \geq 0.5 \end{cases}$$
$$y_r = \begin{cases} K_D - (S \times y_i), & \text{if } N_O - \lfloor N_O \rfloor < 0.5 \\ K_D + 1 - (S \times y_i), & \text{if } N_O - \lfloor N_O \rfloor \geq 0.5 \end{cases} . \quad (4)$$

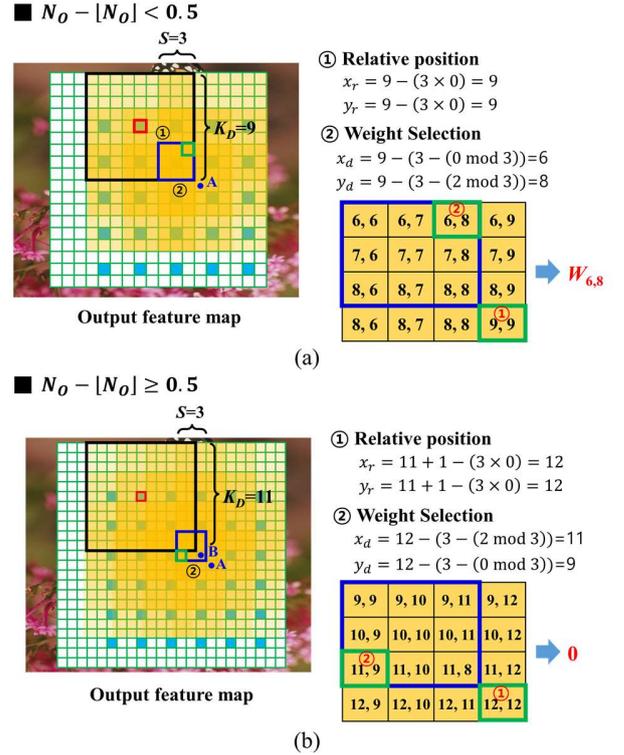

Fig. 7. The process of the inverse coefficient mapping. The relative position depends on (a) when the fractional value of $N_O$ is less than 0.5 and (b) greater than 0.5. During the weight selection, the weight coefficient is taken from $(x_d, y_d)$. However, if $(x_d, y_d)$ exceeds the range of the index, the weight coefficient becomes zero.

Next, we subtract the offset to select the weight coefficient for one of the output pixels as shown in Fig. 7. We calculate the indices of the weight coefficient corresponding to the input pixel as

$$x_d = x_r - (S - (x_o \bmod S))$$
$$y_d = y_r - (S - (y_o \bmod S)) . \quad (5)$$

Finally, $\mathbf{W_D}$, which represents the weights of the deconvolutional layer, is mapped to the weights of the newly created convolutional layer $\mathbf{W_C}$ using Eq. (4) and Eq. (5) as follows.

$$\mathbf{W_C}[S^2 \times m + S \times y_o + x_o][n][y_i][x_i] = \mathbf{W_D}[m][n][y_d][x_d], \quad (6)$$

where $m$ and $n$ are indices for loops of the output and input feature maps with ranges of $1 \leq m \leq M$ and $1 \leq n \leq N$, respectively.

However, if $(x_d, y_d)$ exceeds the range of the indices based on Eq. (3), the weight coefficient becomes zero, thereby producing a zero-valued element, which will be explained in the next sub-section. Therefore, we show that $M \times N$ deconvolution filters are classified as $S^2$ kinds of convolution filters according to the indices of output pixels, as shown in Fig. 8(a). Since $S$ is set to 2 in Fig. 8(a), there are four kinds of sparse convolution filters.

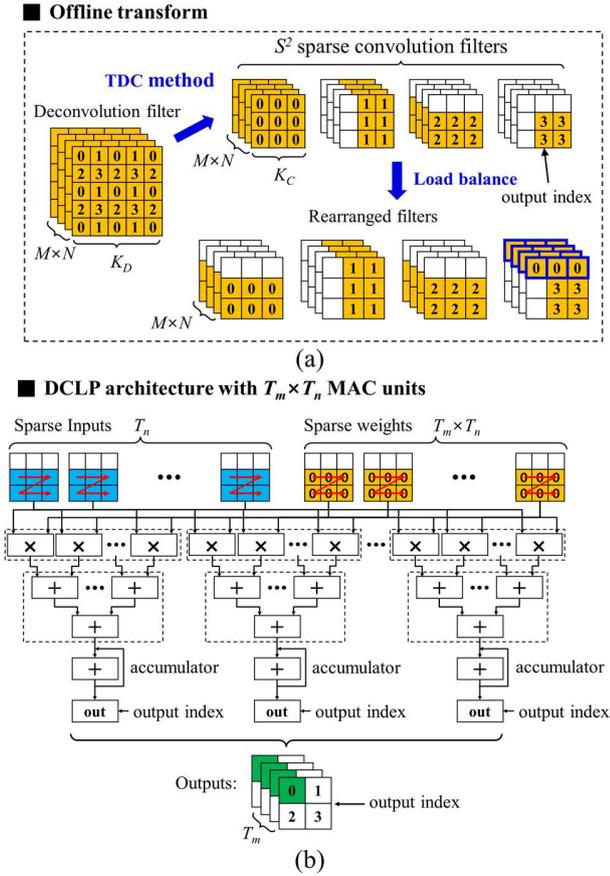

TABLE III
ZERO-WEIGHT RATIO OF CONVOLUTIONAL LAYER CREATED
BY THE TDC METHOD

| $K_D$ | $S$ | $K_C$ | Zero-Weights [%] |
|---|---|---|---|
| 9 | 2 | 5 | 19 |
| 9 | 3 | 3 | 0 |
| 9 | 4 | 3 | 43.8 |
| 7 | 2 | 4 | 23.4 |
| 7 | 3 | 3 | 39.5 |
| 7 | 4 | 2 | 23.4 |
| 5 | 2 | 3 | 30.6 |
| 5 | 3 | 2 | 30.6 |
| 5 | 4 | 2 | 60.9 |

Fig. 8. Dataflow of the proposed deconvolutional layer processor (DCLP). (a) Deconvolution filter is converted to $S^2$ sparse convolution filters offline through TDC method and some of the weights are repositioned for load balance. (b) DCLP architecture with sparse input activations and weights. The results of the rearranged weights are added to the output of the correct position via the output index. In this figure, $K_D$ and $S$ are 5 and 2, respectively, and $K_C$ is set to 3 through the TDC method. Green blocks are output buffers accessed through the output index.

### C. Zero-Aware Processing Element

Our TDC method maps one deconvolution filter with size $K_D \times K_D$ to $S^2$ convolution filters with size $K_C \times K_C$ in each input and output feature map according to Eq. (6). However, they have different sizes because some weights of convolution filters are filled with zero-valued elements. The total number of zero-valued elements, $num_{zero}$, in the transformed convolution kernels is derived as

$$num_{zero} = (K_C^2 \times S^2 - K_D^2) \times M \times N. \quad (7)$$

Table III shows the ratio of zero-weights in the convolutional layer generated by the TDC method. The ratio varies according to $K_D$ and $S$. Moreover, it can be seen that $K_C$ obtained from the TDC method is always smaller than $K_D$.

The TDC method efficiently creates output pixels by reducing the kernel size of the weights. However, the load imbalance problem occurs because of the zero-weights, which is demonstrated by Table III. This is because the distribution of $W_C$ is different for each output pixel.

Fig. 8 shows the dataflow of the proposed DCLP. Before the DCNN inference, we first convert the deconvolution filter to

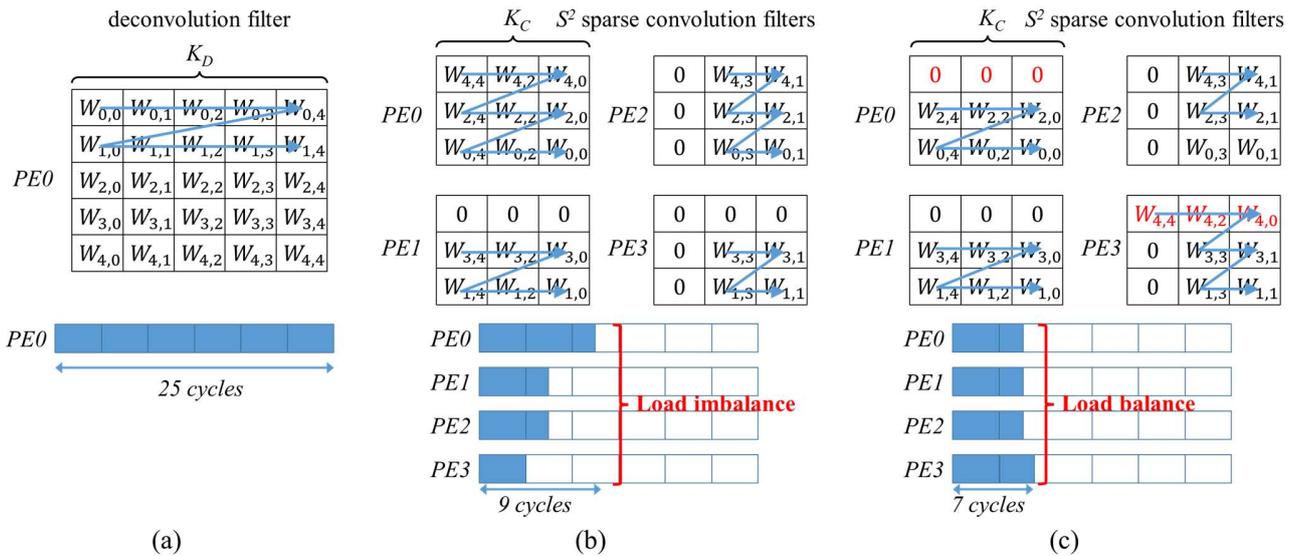

Fig. 9. Advantages of parallelism when TDC and load balance methods are applied. (a) Conventional deconvolutional neural networks (DCNN) accelerator [27]. (b) TDC-based DCNN accelerator in our previous work [28]. (c) Proposed load balance-aware TDC-based DCNN accelerator. In this figure, $K_D$ and $S$ are 5 and 2, respectively, and $K_C$ is set to 3 through the TDC method. $W$ refers to the weight coefficient within each filter and $S$ is the stride.

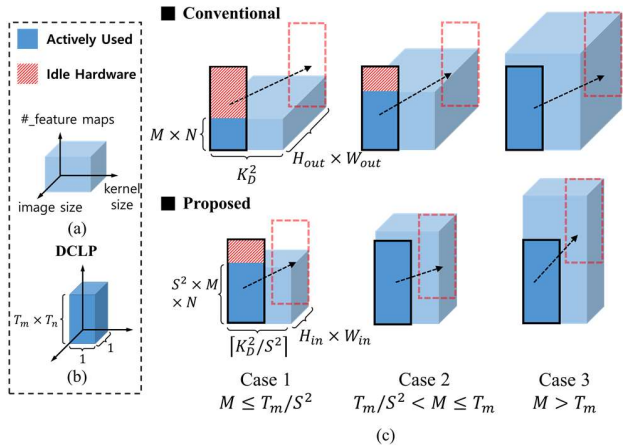

Fig. 10. Comparison of the execution process between the proposed DCLP and the conventional DCLP. (a) Three-dimensional visualization of computational complexity for deconvolutional layer. (b) The DCLP that parallelizes the output feature maps and the input feature maps with tiling parameters $T_m$ and $T_n$. (c) The difference between the proposed method and the conventional method depending on the range of $M$.

convolution filters offline using the TDC method, as shown in Fig. 8(a). We apply a load balance to adjust the proportion of zero weights within each filter. To output the same result as before, we store the output index, the address of the output buffer, in memory along with the weights. We conduct all of the above steps offline. Next, we minimize the execution cycles by evenly distributing non-zero weights across the PEs. Therefore, the total idle cycles of different PEs are reduced. Fig. 8(b) shows the DCLP architecture with sparse input activations and weights. Since the positions of the zero-weights in all the filters are always the same, the positions of the zero inputs are also determined at the same position. As a result, our proposed DCLP exploits both input activation and weight sparsity to balance the load. Finally, we parallelize the input and output feature maps using the loop optimization techniques to make a comparison with the conventional DCNN accelerator in the same environment. DCLP performs MAC operations $T_m \times T_n$ times with inputs and weights. ($T_m$ and $T_n$ are tile sizes for the number of output/input feature maps to parallelize the process)

Until the operation of each filter is completed, the intermediate outputs should accumulate in the previous outputs stored in the buffers. Hence, the results of the reordered weights accumulate in the output buffers via the output index, as shown in Fig. 8(b).

Fig. 9 shows the performance benefits achieved through the proposed load balance-aware TDC method over other methods.

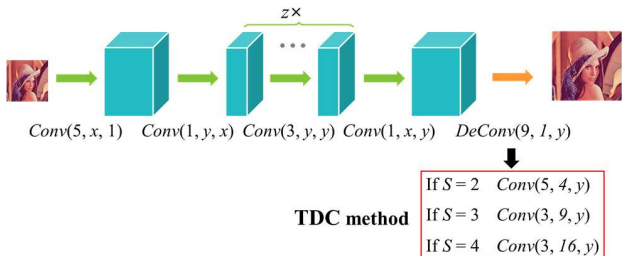

Fig. 11. FSRCNN network structure. For simplicity, we express FSRCNN as $FSRCNN(x, y, z)$, a combination of sensitive variables $x$, $y$, and $z$. (In [25], $x$, $y$, and $z$ are set to 56, 12, and 4, respectively.)

In this case ($K_D$=9, $K_C$=5, $S$=2), we used four PEs. Fig. 9(a) shows the performance of the conventional DCNN accelerator. Fig. 9(b) shows the performance degradation of each PE caused by the load imbalance in our previous work. To efficiently process MAC operation in parallel, we propose the load balance-aware TDC method as shown in Fig. 9(c). In Fig. 9(b), *PE0* contains nine non-zero weights, whereas *PE3* contains four non-zero weights. The pipeline stage is determined by *PE0*, which has the most computational complexity. However, the position of zero-weights is always the same for each deconvolutional layer because the inverse coefficient mapping applies equally to all kernels. Therefore, we design the accelerator to perform load balancing offline, as in Fig. 8(a). The execution cycles of the deconvolutional layer are

Execution cycles
$$= \left\lceil \frac{S^2 \times M}{T_m} \right\rceil \times \left\lceil \frac{N}{T_n} \right\rceil \times \frac{H_{out}}{S} \times \frac{W_{out}}{S} \times \left\lceil \frac{K_D^2}{S^2} \right\rceil . \quad (8)$$
$$= \left\lceil \frac{S^2 \times M}{T_m} \right\rceil \times \left\lceil \frac{N}{T_n} \right\rceil \times H_{in} \times W_{in} \times \left\lceil \frac{K_D^2}{S^2} \right\rceil$$

As shown in Fig. 5, the convolutional layer created using the TDC method produces output feature maps of the same size as the input feature maps, but the number of output feature maps increases by $S^2$ times. The increased number of output feature maps and the input feature maps are processed in parallel by $T_m$ and $T_n$, respectively. Therefore, as opposed to the conventional DCNN accelerator, there are three different cases of performance enhancement depending on the range of $M$; these are demonstrated in Fig. 10. A visualization of the total computational complexity at the deconvolutional layer is shown in Fig. 10(a). Fig. 10(b) shows the hardware size for the DCLP. In the rectangular parallelepiped, the width and height of the bottom surface represent the size of the tiling parameters for the kernel and image, respectively. Because the DCLP performs parallel processing on the input and output feature maps, both lengths are set to 1. Fig. 10(c) shows a visualization of the difference in the performance of the conventional and the proposed method. Both methods are executed on the same DCLP. The three abovementioned cases that are dependent on $M$ are as follows.

Case 1. $M \leq T_m/S^2$

Our method unrolls entire loops for the output feature maps. In addition, we improve the resource underutilization problem, where idle hardware exists in the DCLP, and reduce the convolution cycle by generating LR instead of HR images. The performance enhancement is $S^2 \times \frac{K_D^2}{\lceil K_D^2/S^2 \rceil}$.

Case 2. $T_m/S^2 < M \leq T_m$

Our method completely solves the resource underutilization problem by activating all $T_m$−$M$ hardware resources that are in the idle state. In this case, the performance enhancement is $\frac{S^2}{\lceil (S^2 \times M)/T_m \rceil} \times \frac{K_D^2}{\lceil K_D^2/S^2 \rceil}$.

Case 3. $M > T_m$

Our method cannot process more output feature maps in parallel than the existing DCNN accelerator. However, the execution speed is higher due to the reduced kernel size. The performance enhancement in this case is $\frac{S^2 \times \lceil M/T_m \rceil}{\lceil (S^2 \times M)/T_m \rceil} \times \frac{K_D^2}{\lceil K_D^2/S^2 \rceil}$.

## IV. Proposed DNN-based Super-Resolution System

In this section, we propose a methodology to achieve an energy-efficient architecture for implementing the state-of-the-art DNN-based SR algorithm, FSRCNN. Fig. 11 shows the network structure of the FSRCNN. We express the convolutional layer and the deconvolutional layer as *Conv($K_C$, M, N)* and *DeConv($K_D$, M, N)*, respectively. Using the TDC method, we regard the deconvolutional layer as a convolutional layer. For example, *DeConv($K_D$, M, N)* is converted to *Conv($K_C$, $S^2 \times M$, N)*. In Fig. 11, the reason for denoting variables such as *x*, *y*, and *z* is that they are sensitive variables that determine overall performance in FSRCNN [25]. For simplicity, we represent FSRCNN models with different sensitive variables as FSRCNN (*x*, *y*, *z*). In the conventional FSRCNN model, *x*, *y*, and *z* are experimentally set to 56, 12, and 4, respectively. PReLU [32] is used as an activation function in the FSRCNN.

### A. Dataflow Optimization with On-Chip Memory

In our system, which does not use off-chip memory, the FPGA receives pixel values through the display driver in the horizontal direction of the frame. If all the convolutional layers are executed by a single CLP, the pixel data coming into the FPGA must be stored in the on-chip memory until the last layer is completely processed. Consequently, several frame buffers may be required, depending on the execution time of the CLP. In order to solve this problem, we process the input pixel data by designing all the convolutional layers to run concurrently through multiple CLPs, as in fusion architectures. In this case, multiple CLPs must be designed such that they can handle on-chip dataflow efficiently. To find the loop tiling parameters for multiple CLPs, we compare the execution cycles of each CLP with the transmission cycles of the pixel data coming from the display driver or CLP.

The computation to transmission ratio is defined as the ratio of the cycles required to perform the CLP of the $l^{th}$ convolutional layer to the cycles required to transmit the pixel data from the display driver or input buffers to the $l^{th}$ CLP. Since each CLP must perform several feature maps at the same time, pixel data is sent to the CLP as much as the tiling factor. The CLP performs 2D convolutions on the feature maps with the received pixel data. If the tile sizes for processing the $l^{th}$ convolutional layer are given by $T_m^l$, $T_n^l$, and $T_k^l$ shown in Table I, the computation to transmission ratio is calculated as

*Computation to Transmission Ratio*

$$= \frac{Execution\ cycles\ of\ l^{th}\ CLP}{total\ number\ of\ transmission\ cycles}$$

$$= \frac{\left\lceil\frac{M^l}{T_m^l}\right\rceil \times \left\lceil\frac{N^l}{T_n^l}\right\rceil \times H_{in}^l \times W_{in}^l \times \left\lceil\frac{K_C^l}{T_k^l}\right\rceil \times \left\lceil\frac{K_C^l}{T_k^l}\right\rceil}{\left\lceil\frac{N^l}{T_n^l}\right\rceil \times H_{in}^l \times W_{in}^l}. \quad (9)$$

$$= \left\lceil\frac{M^l}{T_m^l}\right\rceil \times \left\lceil\frac{K_C^l}{T_k^l}\right\rceil \times \left\lceil\frac{K_C^l}{T_k^l}\right\rceil$$

If the computation to transmission ratio is greater than 1, the number of transmission cycles in the $l^{th}$ layer is lower than the number of execution cycles in the CLP. Hence, the data transferred during the computation must be stored in the frame buffer. For example, if an output image with UHD is generated using an SR algorithm with a scale factor of 2, an approximately 8.1 MB buffer memory is required to store an input image with a 1920×1080 resolution in the 32-bit floating-point data type. Furthermore, considering the size of the input feature maps, the required memory can exceed the allowable on-chip memory of a typical FPGA. For this reason, we set the computation to transmission ratio of all layers to a value of 1 in order not to use the frame buffer. Hence, Eq. (9) is equal to 1, and $T_m^l$ and $T_k^l$ become $M^l$ and $K_C^l$, respectively. In addition, the tiling factor of the input feature maps of the next layer ($T_n^{l+1}$) should be the same as that of the output feature maps of the current layer ($T_m^l$); this ensures that there is no buffering. Since $N^{l+1}$ is equal to $M^l$, $T_n^l$ becomes equal to $N^l$. As a result, all the CLPs must parallelize three convolution loops.

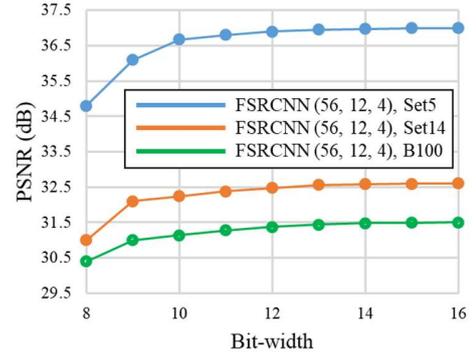

Fig. 12. Bit-width vs. peak signal-to-noise ratio (PSNR) of FSRCNN with fixed-point quantization technique when the up-scaling factor is set to 2.

A memory management technique is required to efficiently store the feature maps generated by multiple CLPs. Since the pixel data from the display driver are transmitted line by line, we use a line buffer that can reuse the data without being restricted by boundary conditions [18]. The line buffer is designed as a block RAM (BRAM), which is a simple dual-port mode [33] in which one read and one write are allowed concurrently in order to fulfill both the input and output buffer of each CLP. The capacity required to implement the line buffer should be same as the size of the 3D data generated from the CLP. The number of line buffers must be equal to the width of the kernel size for the convolution. The size of the line buffers for each CLP is calculated as $K_c^l \times W_{in}^l \times N^l \times$ bit-width.

FSRCNN has two 1×1 convolutional layers as shown in Fig. 11. We unroll all three convolution loops to avoid frame buffering, and therefore, the outputs of the neurons become the inputs of the next neurons without the need to accumulate with the outputs of other neurons. In this manner, the output feature maps generated by the convolutional layer in front of the 1×1 convolutional layer can be directly sent to the CLP of the 1×1 convolutional layer. Thus, we connect the CLP of the layer ahead of the 1×1 convolutional layer and the CLP of the 1×1 convolutional layer without the line buffer. We call this type of processor a combined CLP.

According to a guideline [33], a 7 series FPGA BRAM-18kb unit can store 512 32-bit words. Hence, the number of BRAMs required to generate a UHD image with FSRCNN (56, 12, 4) is 1609. Although the line buffer for the CLP in the 1×1 convolutional layer is removed, it is larger than that used in a typical FPGA. Thus, it is necessary to reduce the usage of BRAMs for the utilization of embedded systems.

*B. Quantization*

Fixed-point implementation mitigates the complexity of hardware design and potentially enables the use of embedded hardware [34]. In particular, our CLPs have three dimensions to process three convolution loops in parallel. Therefore, the implementation of hardware with a fixed-point increases resource utilization efficiency. Fig. 12 shows PSNR according to data bit-width in the representative datasets, Set5, Set14, and B100. Fig. 12 shows that the PSNR decreases dramatically when the bit-width is smaller than 13-bit, while the performance is maintained when the bit-width is larger than 13-bit. This is because the mantissa expressing the fractional value is sufficiently accurate even at low bit-width [35]. To optimize the utilization of the FPGA resources more compactly, pixels, weights, and partial sums were reduced from 32-bit floating-points to 13-bit fixed-points using the bit-width quantization technique [35]. By quantizing the bit-width, we reduced the number of BRAMs from 1609 to 654 when implementing FSRCNN (56, 12, 4).

*C. Compression*

In Xilinx FPGAs, a DSP48E1 block [36] performs up to 25×18-bit multiplication. We must design a 13×13-bit multiplier for convolution on low bit-width data. As a result, there is a problem that resources in the DSP are not sufficiently activated. To solve this problem, we use a double MAC [37], which performs two multiplications on a common operand with a single DSP. The maximum possible bit-width for each operand is 8-bit. We split the 13×13-bit multiplication into 8×8-bit multiplication, 13×5-bit multiplication, and 5×8-bit multiplication and then sum the results. Because CNN iteratively executes multiple operations on the same input feature map, double MAC improves the efficiency of DSP usage. The resource requirement for two 13×13-bit multiplication is 1 DSP+124 LUTs+124 FFs. Due to the high logic element usage of the double MAC, we also considered designing a multiplier with a single DSP called single MAC. The total number of DSPs required in the design of the multiple CLPs can be obtained by

$$\#\_DSPs = \alpha \times \frac{\#\_Multiply}{2} + (1-\alpha) \times \#\_Multiply,$$

where  (10)

$$\#\_Multiply = \sum_{l=1}^{L} M^l \times N^l \times K_C^l \times K_C^l - num_{zero}.$$

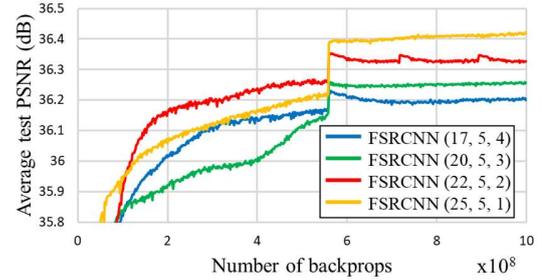

Fig. 13. Convergence curves of different FSRCNN ($x$, $y$, $z$). In this figure, the sensitive variable $y$ is fixed.

$\alpha$ is a parameter that determines the ratio of double MAC. A large value of $\alpha$ can increase image quality, but it lowers power efficiency because a large amount of resources is required [38]. Thus, we experimentally set $\alpha$ to 0.7 considering this trade-off. $L$ is the total number of layers.

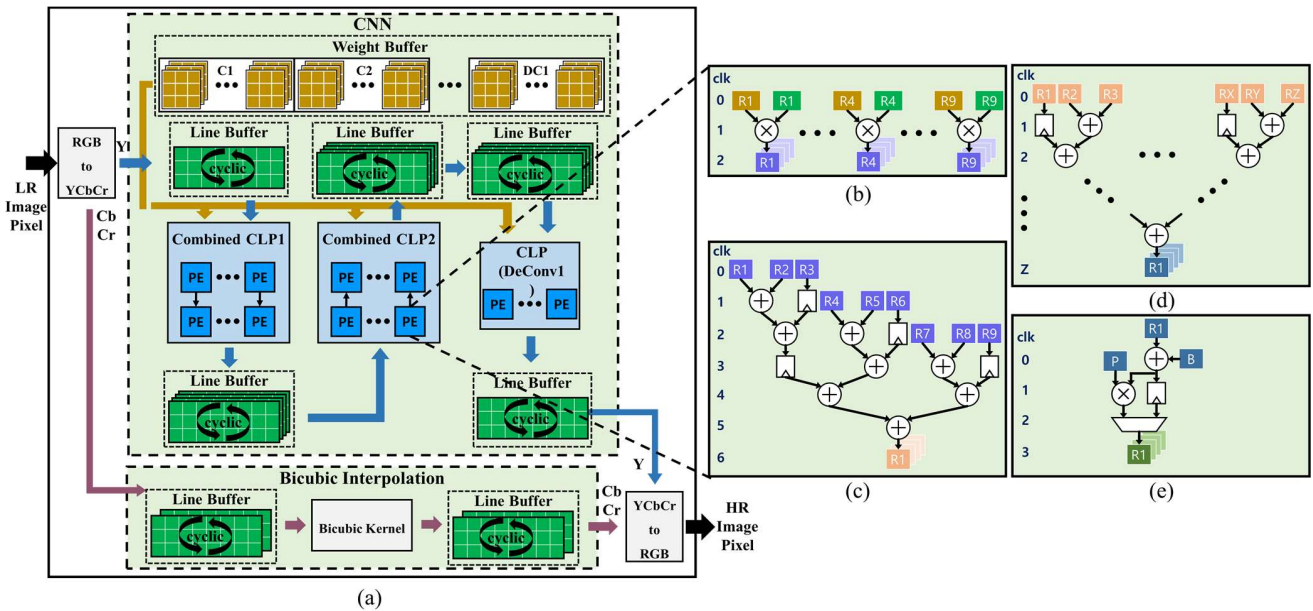

Fig. 14. (a) Overview of the proposed architecture. The processing element of the third convolutional layer consists of (b) multiply engine for kernel-sized convolution, (c) add engine for kernel-sized convolution, (d) add engine for adding all of the input feature maps to generate each output feature map, and (e) activation engine for generating output of neuron.

TABLE IV
COMPARISON OF PSNR (SET5) AND RESOURCE UTILIZATION REQUIRED FOR DIFFERENT SETTINGS (FT: FLOATING-POINT, FX: FIXED-POINT)

| x | y | z | DSP | BRAM | PSNR (FT) | PSNR (FX) |
|---|---|---|---|---|---|---|
| 17 | 5 |   | 1,514 (98%) | 194 (24%) | 36.21 | 36.17 |
| 21 | 4 | 4 | 1,494 (97%) | 205 (26%) | 36.23 | 36.18 |
| 25 | 3 |   | 1,511 (98%) | 215 (27%) | 36.22 | 36.17 |
| 20 | 5 |   | 1,531 (99%) | 194 (24%) | 36.27 | 36.23 |
| 23 | 4 | 3 | 1,507 (98%) | 202 (25%) | 36.29 | 36.24 |
| 26 | 3 |   | 1,510 (98%) | 210 (26%) | 36.24 | 36.20 |
| 22 | 5 |   | 1,497 (97%) | 188 (24%) | 36.36 | 36.32 |
| 24 | 4 | 2 | 1,482 (96%) | 193 (24%) | 36.42 | 36.38 |
| 26 | 3 |   | 1,492 (97%) | 198 (25%) | 36.30 | 36.26 |
| 25 | 5 |   | 1,512 (98%) | 188 (24%) | 36.43 | 36.40 |
| 26 | 4 | 1 | 1,480 (96%) | 191 (24%) | 36.37 | 36.33 |
| 28 | 3 |   | 1,509 (98%) | 200 (25%) | 36.29 | 36.26 |

TABLE V
PERFORMANCE COMPARISON OF DCNN ACCELERATORS

| Model | Layer | $K_D$ | $K_C$ | S | $T_m$ | $T_n$ | [28] Cycles×1000 | Ours Cycles×1000 |
|---|---|---|---|---|---|---|---|---|
|       | 1 | 5 | 3 | 2 | 4 | 128 | 1,638 | 458 |
|       | 2 | 5 | 3 | 2 | 4 | 128 | 1,638 | 458 |
| DCGAN | 3 | 5 | 3 | 2 | 4 | 128 | 1,638 | 458 |
|       | 4 | 5 | 3 | 2 | 4 | 128 | 102 | 21 |
|       | Total |   |   |   |   |   | 5,017 | 1,397 |
|       |   | 8 | 9 | 5 | 2 | 56 | 9 | 21,233 | 1,376 |
| FSRCNN |  | 8 | 9 | 3 | 3 | 56 | 9 | 47,775 | 589 |
|       |   | 8 | 9 | 3 | 4 | 56 | 9 | 84,934 | 786 |

The number of DSPs required to implement FSRCNN (56, 12, 4) is 8,102 when α=0.7. This requirement is higher than the total number of DSPs embedded in high-end FPGAs. For this reason, we compress the FSRCNN into an optimal model for efficient inference.

We change the sensitive variables of FSRCNN considering the total number of DSPs in the target FPGA specification. We set the target FPGA to the Kintex-7 410T FPGA, which is a cost-effective digital processing platform. Since the number of parameters used in the deconvolutional layer is approximately 50% of the total due to large kernel size, we reduce the kernel size of the deconvolutional layer from 9×9 to 7×7 so that sensitive parameters can take larger values. Table IV shows the average PSNR on the Set5 dataset and resource usage for various sensitive variables when the scale factor is 2. We train the models offline via the GPU with the Caffe framework [39]. Due to a lack of hardware resources, sensitive variables cannot be large values. When the number of parameters is fixed, we see that performance is better when z is smaller. This trend also appears in the convergence curves shown in Fig. 13. This is because if z is large, x is too small for the convolutional layer to extract enough local features for reconstruction.

As shown in Table IV, the model with the highest PSNR is FSRCNN (25, 5, 1). This model is also resource efficient because it has the lowest BRAM usage. As a result, we implement FSRCNN (25, 5, 1), which is a light version of FSRCNN, in hardware.

### D. Hardware Implementation

Fig. 14(a) shows the proposed on-chip memory-based FPGA architecture. The FPGA receives an input LR image with RGB channels and uses the Y channel after performing RGB-to-YCbCr conversion. In general, in DNN-based SR systems, the Cb and Cr channels are rarely used for learning [25]. Thus, we up-scaled the Cb and Cr channels with bicubic interpolation.

Before entering the CLP, the pixel data are stored in the line buffers. After $K_C^l$-1 lines are stored, the outputs of the line buffers and incoming data enter the CLP; then convolution is performed with the filters from the weight buffer. Fig. 14(a) shows that the CLPs of the first layer and the second layer are fused into the combined CLP1. In addition, the CLPs of the third layer and the fourth layer are fused into the combined CLP2. By using the combined CLPs, we could reduce the total amount of line buffers to 83%.

Figs. 14(b)–(e) show the computation engines in the PE constituting the CLP of the third layer. First, the PE fetches the data stored in the line buffer into the registers and performs multiplication with weights. We perform all the multiplication operations within the kernel at the same time and process these operations $M^l×N^l$ times in parallel as depicted in Fig. 14(b). Then, we add the outputs of the multiplication through the adder tree, as shown in Fig. 14(c). Fig. 14(d) shows the process of adding the results of each input feature map through the add engine consisting of $M^l$ adder trees. Finally, Fig. 14(e) shows the output of the neurons via the PReLU activation engine.

## V. EXPERIMENTAL RESULTS

### A. Evaluation of the Proposed DCNN Accelerator

We validated the DCNN accelerators with the Xilinx Virtex-7 485T FPGA in the same experimental environment as that used by the study in [28]. We implemented the proposed and the conventional architecture using Vivado HLS 2016.4 and a single-precision floating-point. We evaluated the DCNN models FSRCNN and DCGAN [40] on the hardware used in previous studies [27], [28]. To compare the performance of the proposed DCNN accelerator in the same experimental environment as the conventional DCNN accelerator, we designed the accelerator using the single CLP method [10]. The conventional DCNN accelerator paralleled the convolution loops for output feature maps and input feature maps with $T_m$ and $T_n$, respectively, and determined the optimal tiling parameters through the roofline model [41]. Fig. 15 shows possible design space solutions when designing the CLP for the fourth layer of the FSRCNN by means of the roofline model. The computation to communication ratio is the number of operations performed per external memory access. Therefore, in order to utilize all possible hardware resources and minimize the bandwidth in off-chip memory communications, we chose the optimal solution for each layer, as depicted in Fig. 15. Then, we performed cross-layer optimization [28]. We set the tiling parameters ($T_m$, $T_n$) for the FSRCNN and DCGAN to (56, 9) and (4, 128), respectively. Table V shows the performance comparison of the existing method and the proposed method. The performance analysis of the proposed accelerator for DCGAN and FSRCNN is as follows.

First, DCGAN consists of four deconvolutional layers. Each layer has a greater number of input feature maps than output feature maps. Thus, $T_m$ was set to 4, which is 42 times smaller than $T_n$. Our load balance-aware TDC method improved performance even further when the resource underutilization problem existed in the conventional accelerators because $M$ was smaller than $T_m$. In DCGAN, this situation occurred in the last deconvolutional layer. However, the speedup was not significant, because there was little difference between $T_m$ and $M$. Since the resource underutilization problem was not apparent in the CLP of the DCGAN, the performance was improved only by the advantage of performing kernel computation in shorter cycles. Therefore, the proposed method was 3.59 times faster than the conventional method in the DCGAN.

TABLE VI
PERFORMANCE COMPARISON WITH STATE-OF-THE-ART CNN AND DCNN ACCELERATORS (FX: FIXED-POINT, FT: FLOATING-POINT)

|  | [12] AlexNet | [20] AlexNet | [13] VGG16 | [16] VGG16 | [20] VGG16 | [21] VGG16 | [17] VGG19 | [18] VGG19 | [27] DCGAN | This Work Light FSRCNN |
|---|---|---|---|---|---|---|---|---|---|---|
| FPGA | Startix-V GXA7 | Startix-V GXA7 | Zynq XC7Z045 | Arria-10 GX1150 | Startix-V GXA7 | Arria-10 GX1150 | Virtex-7 XC7V690T | Zynq XC7Z045 | Zynq XC7Z020 | Kintex-7 XC7K410T |
| Network | CNN | CNN | CNN | CNN | CNN | CNN | CNN | CNN | DCNN | CNN+DCNN |
| Output Resolution | 1×1×1000 | 1×1×1000 | 1×1×1000 | 1×1×1000 | 1×1×1000 | 1×1×1000 | 1×1×1000 | 1×1×1000 | 64×64×3 | 2880×1280×3 (QHD) |
| Frequency | 100 MHz | 200 MHz | 150 MHz | 150 MHz | 200 MHz | 385 MHz | 100 MHz | 100 MHz | 100 MHz | 130 MHz |
| Precision | 8-16 bit FT | 32 bit FT | 16 bit FX | 8-16 bit FX | 32 bit FT | 16 bit FX | 32 bit FT | 16 bit FX | 12 bit FX | 13 bit FX |
| DSP Usage | 256 | 224 | 780 | 1,518 | 224 | 1,378 | 784 | 824 | 220 | 1,512 |
| LUTs Usage | 112K | 200K | 183K | 161K | 200K | - | 118K | 155K | 25K | 167K |
| FFs Usage | - | 266K | 128K | - | 266K | - | 90K | 120K | 30K | 158K |
| Memory Size | 3.9MBytes | 4.0MBytes | 2.2MBytes | 4.8MBytes | 4.0MBytes | 3.6MBytes | 3.2MBytes | 4.1MBytes | 0.3MBytes | 0.9MBytes |
| $Cycles(\times 10^3)$ | 1,161.6 | 3,204.8 | 33,886.9 | 7,194.9 | 50,121.5 | 6,656.8 | 24,059.2 | 16,993.4 | 65,384.6 | 2,073.6 |
| Power | 12.9W | 13.2W | 9.6W | 21.2W | 13.2W | 37.4W | 9.4W | 9.4W | - | 5.4W |
| Throughput (GOPS) | 114.5 | 83.0 | 137.0 | 645.3 | 123.5 | 1,790 | 162.1 | 229.5 | 2.6 | $S$=2 780 / $S$=3 1,576.3 / $S$=4 2691 |
| Power Efficiency (GOPS/W) | 8.9 | 6.3 | 14.2 | 30.4 | 9.37 | 47.8 | 17.2 | 24.4 | - | $S$=2 144.9 / $S$=3 293.0 / $S$=4 500.2 |

Secondly, FSRCNN uses the deconvolutional layer as the last layer and can set the resolution of the output image according to $S$. The width of the kernel size $K_D$ is 9, as shown in Table V. Unlike in DCGAN, 88.9% of the hardware resources were idle in the deconvolutional layer of FSRCNN. This is because the number of output feature maps $M$ was nine times smaller than $T_m$. Thus, our load balance-aware TDC method could reduce the ratio of idle hardware to 55.5% when the value of $S$ was 2. Even when the value of $S$ was 3, all the idle hardware was activated, resulting in a performance improvement of 81 times compared to the conventional method. However, as shown in Table III, there were no zero-valued weights in the filters, and therefore we could not take advantage of sparse matrix multiplication in this case. However, Table III shows that the ratio of zero-valued weights was 43.8% when the value of $S$ was 4. In other words, the load imbalance was serious, because the difference in the activated resources of the PEs was large. As a result, we evenly distributed the operations that the PEs executed unequally in conventional kernel computation. Therefore, our accelerator was able to run 108 times faster than the conventional accelerator using the same hardware resources.

### B. Implementation Results of Proposed SR System

We evaluated our proposed DNN-based SR system using Vivado 2016.4. We designed our overall architecture with Verilog RTL. Our FPGA was connected to a 2880×1280 (QHD) panel.

Table VI shows a comparison of the implementation specifications of our accelerator with various FPGA-based accelerators. We implemented Light FSRCNN, which reduced the model size of the FSRCNN. Since we compressed the model size to maximize the utilization of the DSP in the Kintex-7 410T FPGA, the DSPs were fully utilized. Additionally, BRAM usage was 26% of the total available, with the advantage that the CNN model was smaller. Fig. 16 shows the hardware resources breakdown result of the Light FSRCNN. Among all the modules, DeConv1 module has the highest resource usage. We simulated using Xilinx Power Estimation and Analysis Tools to measure the power consumption of the FPGA board. The total thermal power was 5.38W consisting of sources from combined CLP1 (20.7%), combined CLP2 (14.3%), DeConv1 (35.3%), I/O (8.2%), controller (4.5%), interface (11.47%), and others (5.4%). DNN modules that used a lot of DSP and logic cells had higher power consumption than other sources.

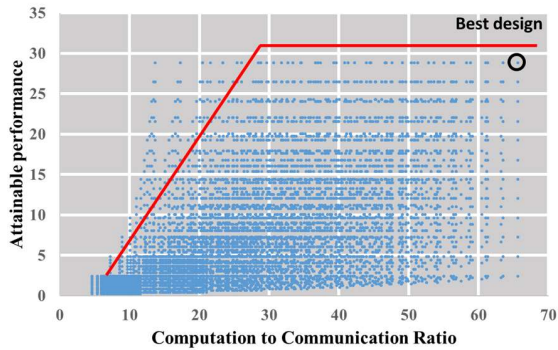

Fig. 15. Example of design space exploration.

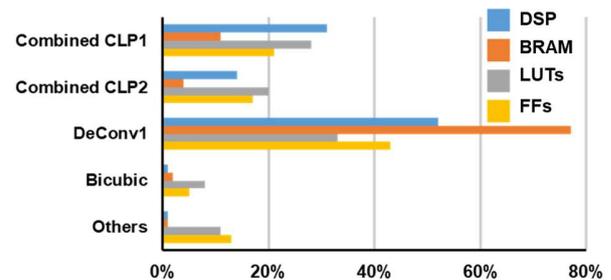

Fig. 16. Hardware resources breakdown when running Light FSRCNN.

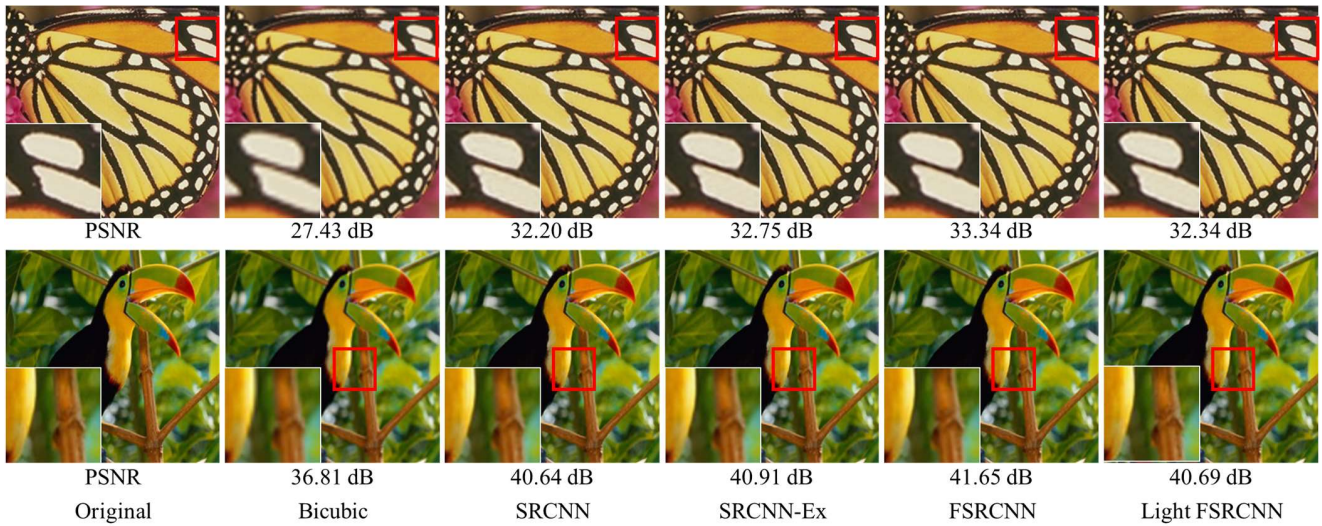

Fig. 17. Performance comparison of SR algorithms when an up-scaling factor is 2 in butterfly and bird images.

In conventional accelerators, the architecture is designed for either CNNs or DCNNs. However, the design of our accelerator is the first that has a hybrid form in which both CNNs and DCNNs are implemented together in the hardware platform. The throughput (GOPS) of each implemented accelerator, shown in the Table VI, was computed as the total computational complexity for spatial convolution divided by the average execution time per image. However, DCNNs increase the computational complexity in proportion to the power of $S$. For this reason, the rate of computational complexity occupied by the deconvolutional layer in Light FSRCNN was 81.67%, 90.92%, and 94.68% when the value of $S$ was 2, 3, and 4, respectively. However, our DCNN accelerator solved the large loop dimension problem of the output image by using the TDC method. This is because we could simultaneously generate HR images with $S^2$ channels of LR images using the TDC method, GOPS was higher in proportion to $S$.

Compared to the other accelerators for object detection and recognition, the OpenCL-based method [21] had the highest throughput using the highest clock rate. However, when power consumption was considered, our accelerator had the highest power efficiency. Our system consumed less power by not using off-chip memory and achieved high performance in Light FSRCNN.

Table VI shows the total execution cycles required to run each DNN. Since the computational complexity of AlexNet is approximately 20 times smaller than VGG16 [5], AlexNet-based accelerator [12] required the fewest cycles. Light FSRCNN has 1.2 times less computational complexity than VGG16. However, our accelerator was at least 3.6 times faster than the other accelerators based on VGG16.

In [20], the authors demonstrated that convolution can be performed in the frequency domain through fast Fourier transformation. Although their method had the advantage of less hardware resource usage in relation to performance, it required high power because it included a CPU together with off-chip memory. Another example of an accelerator, the state-of-the-art fusion architecture [18], reduced the amount of off-chip data transfer more than the conventional fusion architecture [17] by optimizing the dataflow between adjacent layers using more BRAMs. As a result, the amount of external memory access was reduced to improve power efficiency. However, a drawback remained in that power was still required for the off-chip memory. Without using off-chip memory, we enhanced power efficiency with an optimized dataflow for on-chip memory. Since we had to move large amounts of intermediate data to BRAMs, FFs usage was higher than other fusion architectures. However, our CNN accelerator was at least three times more power efficient than other accelerators.

Table VII shows the hardware implementation results of our proposed system as compared to those of existing SR systems. Yang et al. [42] implemented anchored neighborhood regression (ANR) in hardware to generate FHD images at 60fps, and Kim et al. [43] generated UHD images at 60 fps using the super-interpolation (SI) method. However, the scale factor supported by their methods was fixed to 2, and therefore, a limitation existed in that they could not generate an output image with a larger resolution image in the same hardware architecture. Our DNN-based SR system required more hardware resources than conventional methods, but can support a variety of scale factors through the deconvolutional layer with the same hardware resources. A virtual input/output (VIO) core [44] is a customizable core that allows virtual inputs and outputs to be added to hardware description language design. This core allowed us to drive internal FPGA signals synchronously or asynchronously. The FSRCNN is characterized by the fact that the weights of the convolutional layers do not change even if the scale factor changes; only the weights of the deconvolutional layer change [25]. We pre-stored the weights of all the deconvolutional layers, each 3.98KB in size, in ROM using the VIO core. Thus, we could obtain the output by adjusting the internal signals through the VIO core without having to re-synthesize to change the weights of the deconvolutional layer stored in the ROM when different scale factors were required.

TABLE VII
COMPARISON OF DIFFERENT SR SYSTEMS (FX: FIXED-POINT)

|  | [42] | [43] | Proposed |
|---|---|---|---|
| Methods | ANR | SI | Light FSRCNN |
| FPGA | Altera EP4S GX530 | Kintex UltraScale XCKU040 | Kintex-7 XC7K410T |
| Frequency | 136 MHz | 150 MHz | 130 MHz |
| Precision | FX | FX | 13 bit FX |
| FPGA Resources | - | LUTs: 3K FFs: 20K DSP Blocks: 108 | LUTs: 167K FFs: 158K DSP Blocks: 1512 |
| Memory Size | 235KBytes | 92KBytes | 945KBytes |
| Power | - | - | 5.38W |
| Implementation | FHD 60fps | UHD 60fps | QHD 141fps |
| Supported Scale | 2X | 2X | 2X, 3X, 4X |

TABLE VIII
IMAGE QUALITY COMPARISON OF VARIOUS METHODS

| | | Bicubic | | [42] ANR | | [43] SI | | [23] SRCNN | | [24] SRCNN-Ex | | [25] FSRCNN | | [25] FSRCNN-s | | Ours Light FSRCNN | |
|---|---|---|---|---|---|---|---|---|---|---|---|---|---|---|---|---|---|
| # of Parameters | | - | | - | | - | | 8,032 | | 57,184 | | 12,464 | | 3,937 | | 2,325 | |
| Bit-width | | - | | - | | 12-bit | | 32-bit | | 32-bit | | 32-bit | | 32-bit | | 13-bit | |
| Test Sets | Scale | PSNR | SSIM | PSNR | SSIM | PSNR | SSIM | PSNR | SSIM | PSNR | SSIM | PSNR | SSIM | PSNR | SSIM | PSNR | SSIM |
| Set5 | 2 | 33.66 | 0.9299 | 33.83 | - | 34.78 | 0.9460 | 36.34 | 0.9521 | 36.66 | 0.9542 | 37.00 | 0.9558 | 36.57 | 0.9532 | 36.40 | 0.9527 |
| Set14 | 2 | 30.23 | 0.8688 | 29.77 | - | 31.63 | 0.9083 | 32.18 | 0.9039 | 32.45 | 0.9067 | 32.63 | 0.9088 | 32.28 | 0.9052 | 32.21 | 0.9047 |
| B100 | 2 | 29.56 | 0.8431 | - | - | 30.49 | 0.8776 | 31.13 | 0.8850 | 31.36 | 0.8879 | 31.50 | 0.8908 | 31.23 | 0.8866 | 31.15 | 0.8858 |
| Set5 | 3 | 30.39 | 0.9299 | - | - | - | - | 32.39 | 0.9033 | 32.75 | 0.9090 | 33.16 | 0.9140 | 32.54 | 0.9055 | 32.48 | 0.9043 |
| Set14 | 3 | 27.54 | 0.7736 | - | - | - | - | 29.00 | 0.8145 | 29.30 | 0.8215 | 29.43 | 0.8242 | 29.08 | 0.8167 | 29.03 | 0.8146 |
| B100 | 3 | 27.21 | 0.6675 | - | - | - | - | 28.21 | 0.7807 | 28.41 | 0.7863 | 28.52 | 0.7900 | 28.33 | 0.7815 | 28.25 | 0.7808 |
| Set5 | 4 | 28.42 | 0.8104 | - | - | - | - | 30.09 | 0.8530 | 30.48 | 0.8628 | 30.71 | 0.8657 | 30.11 | 0.8499 | 30.17 | 0.8532 |
| Set14 | 4 | 26.00 | 0.7019 | - | - | - | - | 27.20 | 0.7413 | 27.49 | 0.7503 | 27.59 | 0.7535 | 27.19 | 0.7423 | 27.24 | 0.7414 |
| B100 | 4 | 25.96 | 0.6675 | - | - | - | - | 26.71 | 0.7035 | 26.90 | 0.7101 | 26.96 | 0.7139 | 26.84 | 0.7085 | 26.71 | 0.7041 |

Table VII also shows that our system could generate QHD at 141 fps when the scale factor is 2. In the case where a UHD video stream was required, our system could generate UHD images at 62.7 fps using approximately twice the number of BRAMs. When the scale factor was greater than 2, the speed of our system was inversely proportional to the input resolution. For example, if the scale factor was 3, an image with a resolution of 1280×720, which is smaller than the FHD, was used as the input to generate the UHD image.

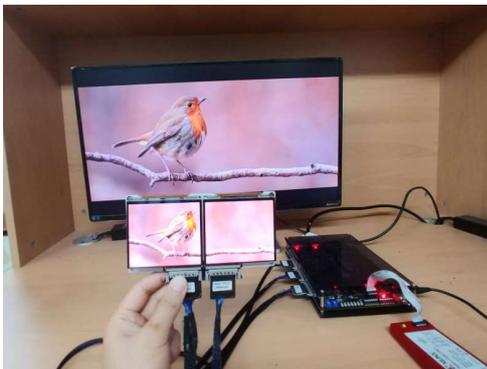

Fig. 18. Demonstration of the DNN-based super-resolution system.

Table VIII shows a comparison in terms of image quality of various SR methods in different scale factors. We evaluated the performance of the SR systems and algorithms on the datasets that are most frequently used. Although our system had a lower performance than the FSRCNN for hardware implementation, our method could achieve higher image quality than existing systems. Fig. 17 shows the reconstructed images of CNN-based SR algorithms and bicubic interpolation. Our Light FSRCNN used 5,707 fewer parameters than SRCNN, but output images had perceptual quality similar to those of other algorithms.

Fig. 18 demonstrates our DNN-based SR system in a mobile panel. We confirmed that HR images can be generated from the QHD panel for mobile applications. In the future, we will further increase the sparsity of the DNN to implement the DNN-based SR as a more hardware-friendly architecture. In addition, we will improve the resource efficiency of our DCNN accelerator by transforming the feature maps into the frequency domain.

## VI. CONCLUSION

In this paper, we proposed an energy-efficient DNN-based SR architecture for hardware implementation. First, we presented a novel methodology to optimize the dataflow for effectively designing the DCNN with higher computational complexity than the CNN in hardware implementation. In addition, we proposed an energy-efficient DNN architecture. Our experimental results showed that our DCNN accelerator achieved a speed of up to 108 times faster than a conventional DCNN accelerator with the same hardware resources. Moreover, the proposed DNN-based SR system was shown to be at least three times more power efficient than the state-of-the-art implementations.

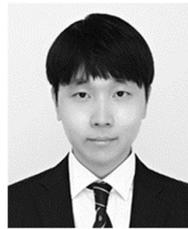

**Jung-Woo Chang** received the B.S degree in electronics engineering from Sogang University, Seoul, South Korea, in 2016, where he is currently pursuing the M.S. degree.

His current research interests include hardware acceleration of deep learning and computer architecture.

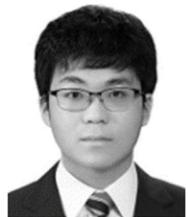

**Keon-Woo Kang** received the B.S degree in electronics engineering from Sogang University, Seoul, South Korea, in 2018, where he is currently pursuing the M.S. degree.

His current research interests include efficient hardware design for deep learning acceleration.

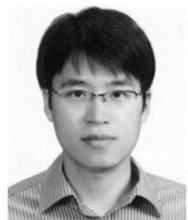

**Suk-Ju Kang** received the B.S. degree in electronics engineering from Sogang University, Seoul, South Korea, in 2006, and the Ph.D. degree in electrical and computer engineering from Pohang University of Science and Technology, Pohang, South Korea, in 2011.

From 2011 to 2012, he was a Senior Researcher with LG Display, Seoul, where he was a Project Leader for resolution enhancement and multiview 3D system projects. From 2012 to 2015, he was an Assistant Professor of Electrical Engineering with Sogang University. His research interests include image analysis and enhancement, video processing, multimedia signal processing, and circuit design for LCD, OLED, and 3D display systems.